\documentclass{emulateapj}

\shorttitle{Revealing the Molecules Ejected from Grains}
\shortauthors{I. Jim\'{e}nez--Serra et al.}

\begin{document}

\title{Grain Evolution across the Shocks in the L1448--mm Outflow}

\author{I. Jim\'{e}nez--Serra, J. Mart\'{\i}n--Pintado and A.
Rodr\'{\i}guez--Franco\altaffilmark{1}}
\affil{Departamento de Astrof\'{\i}sica Molecular e Infrarroja, 
Instituto de Estructura de la Materia,
Consejo Superior de Investigaciones Cient\'{\i}ficas (CSIC),
C/ Serrano 121, E--28006 Madrid, Spain; 
izaskun@damir.iem.csic.es, martin@damir.iem.csic.es,
arturo@damir.iem.csic.es}

\and

\author{S. Mart\'{\i}n}
\affil{Instituto de Radioastronom\'{\i}a Milim\'etrica,
Avda. Divina Pastora, Local 20,
E--18012 Granada, Spain; martin@iram.es}

\altaffiltext{1}{Escuela Universitaria de \'Optica,  
Departamento de Matem\'atica Aplicada (Biomatem\'atica),
Universidad Complutense de Madrid,
Avda. Arcos de Jal\'on s/n, E--28037 Madrid, Spain}

\begin{abstract}

The recent detection of the shock--precursors toward the very young
L1448--mm outflow offers the possibility to study the grain chemistry
during the first stages of the shock evolution, constraining the
molecules ejected from grains and the species formed in gas phase.  
Observations of key molecules in the grain chemistry
like SiO, CH$_3$OH, SO, CS, H$_2$S, OCS, and SO$_2$ toward this
outflow are presented. The line
profiles and the derived abundances show three distinct velocity
regimes that trace the shock evolution: the preshock, the
shock--precursor and the postshock gas. The SiO, CH$_3$OH, SO, and CS
abundances are enhanced with respect to the quiescent gas 
by one order of magnitude in the shock--precursor component, 
and by three orders of magnitude in the postshock gas. 
The derived SiO and CH$_3$OH abundances are
consistent with the recent ejection of these molecules from grains.
Since H$_2$S is only enhanced in the shock--precursor component, and
OCS and SO$_2$ are undetected, SO and
CS are the most abundant sulfur--bearing species in the grain
mantles of L1448--mm. The ejection of mainly SO and CS rather than
H$_2$S or OCS from grains, suggests that the sulfur chemistry will depend on 
the chemical ``history'' of the grain mantles in outflows and hot
cores.

\end{abstract}


\section{Introduction}

Observational studies of the abundances of shock tracers
like SiO, CH$_3$OH and the sulfur--bearing molecules H$_2$S, SO, CS, 
OCS or SO$_2$ toward relatively evolved (t$\geq$10$^{4}$$\,$yr) 
outflows and hot cores (Blake et al. 1987; Bachiller \&
P\'erez--Guti\'errez 1997; van der Tak et al. 2003), 
have established that these molecules 
are largely enhanced ($\sim$10$^{-7}$) 
due to the release of material from grains. 
Since silicon is heavily depleted onto grains 
(Ziurys, Friberg, \& Irvine 1989; Mart\'{\i}n--Pintado, 
Bachiller, \& Fuente 1992) and CH$_3$OH has been firmly detected on icy 
mantles (Tielens \& Allamandola 1987), SiO and CH$_3$OH are accepted
to be ejected from grains. For the sulfur chemistry, 
since H$_2$S is predicted to be efficiently formed 
on grain surfaces (Duley, Millar, \& Williams 1980), this molecule is
assumed to initiate this chemistry (Charnley 1997). 
However, H$_2$S has not been detected so far in solid 
state \citep{smi91}. Although other species like OCS have been
proposed to play an important role in the sulfur chemistry
\citep{vdt03,smar04}, the question of which sulfur--bearing molecule 
is ejected from grains still remains uncertain.

Shock chemistry models that include injection of molecules from
grains, show that the abundances of key species like SiO, CH$_3$OH or
H$_2$S evolve in time--scales of few 10$^3$$\,$yr (Charnley et
al. 1992; Flower et al. 1996; Markwick, Millar, \& Charnley 2000), 
much shorter than the typical ages of evolved outflows.
The study of the chemistry of very young outflows like L1448--mm with
dynamical time--scales (t$\sim$3500$\,$yr) similar to the chemical 
time--scales, offers the possibility to establish which molecules are 
formed or depleted onto grains and which ones are generated in gas phase. 
In particular, since magnetohydrodynamic (MHD)
C--shocks are initiated by the magnetic precursor,
one can directly measure the material released from grains by
observing the gas recently affected by the precursor. 
The detection of extremely narrow SiO emission toward 
L1448--mm, and the increase of the molecular ion abundances, have been 
interpreted as signatures of the interaction of the precursor 
\citep{jim04}. Measurements of the
abundances of shock tracers in the shock--precursor
stage can therefore help to establish the molecules that are
present on grain mantles. This is particularly important in the case of
the sulfur--chemistry.

In this Letter, we present observations of key shock tracers
like SiO, CH$_3$OH, SO, CS, H$_2$S, OCS and SO$_2$, 
toward the L1448--mm outflow. The enhanced abundances 
of SiO, CH$_3$OH, SO, and CS 
in the shock--precursor and the postshock components indicate
the recent ejection of these molecules from grains. This enhancement,
the lack of SO$_2$ and OCS, and the fact that
H$_2$S is only enhanced in the shock--precursor component, 
indicate that the sulfur chemistry in L1448--mm strongly 
depends on the evolution of the chemical composition of grains.  

\begin{figure*}
\epsscale{0.94}
\plotone{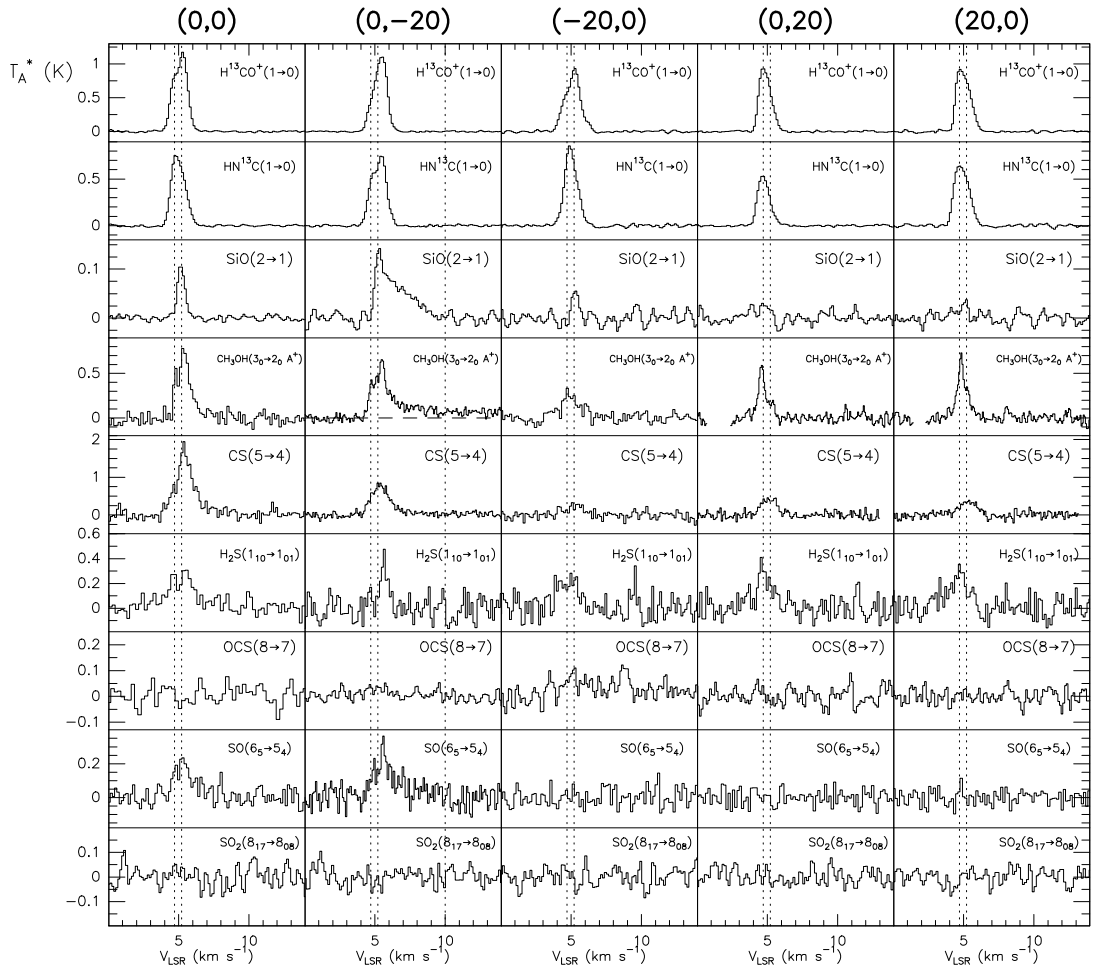}
\caption{Sample of the molecular profiles of H$^{13}$CO$^{+}$,
  HN$^{13}$C, SiO, CH$_3$OH, CS, H$_2$S, OCS, SO, and SO$_2$ 
  observed toward the 5--point map of L1448--mm. 
  Offsets in arcseconds relative to L1448--mm
  ($\alpha$(2000)~=~03$^{h}$25$^{m}$38$_.^s$0, 
  $\delta$(2000)~=~30$^{\circ}$44$'$05$''$) 
  are shown in the upper part of the columns.
  The dotted vertical lines show the preshock component at 
  4.7$\,$km$\,$s$^{-1}$ and the shock--precursor component at
  5.2$\,$km$\,$s$^{-1}$. Toward L1448--mm (0,--20), SiO, CH$_3$OH, CS,
  and SO show broader profiles with terminal velocities that vary from
  7.5$\,$km$\,$s$^{-1}$ for CS, to 14$\,$km$\,$s$^{-1}$ for CH$_3$OH
  (see dashed horizontal line). SiO and SO have similar terminal velocities
  of $\sim$10$\,$km$\,$s$^{-1}$ (third dotted line).}
\label{fig1}
\end{figure*}

\section{Observations \& Results}

We mapped the L1448--mm outflow observing toward the 5--point map
previously measured in SiO, H$^{13}$CO$^{+}$, and HN$^{13}$C by 
Jim\'enez--Serra et al. (2004). The observations were carried out with 
the IRAM 30$\,$m telescope at Pico Veleta (Spain).
The beam size was $\sim$27$''$, 16$''$, and 10$''$ at
$\sim$90, 150, and 240$\,$GHz. The SIS receivers were tuned to
single side band with image rejections $\geq$10$\,$dB. 
We observed lines of SO$_2$, H$_2$S, SO, OCS, CH$_3$OH, and CS.
Observations were done in frequency switching and wobbler 
switching modes with frequency and position throws of
7.2$\,$MHz and 240$''$ respectively. A spectral resolution of
$\sim$40$\,$kHz was achieved with the autocorrelators, 
which corresponds to velocity resolutions of $\sim$0.14, 0.08, and
0.05$\,$km$\,$s$^{-1}$ at the observed frequencies. 
The system temperatures were typically
$\sim$150--400$\,$K. Toward L1448--mm (0,--20), we
also observed the CO \textit{J}=1$\rightarrow$0 transition in
position switching mode with the 1MHz filter bank, that provided 
a velocity resolution of $\sim$2.6$\,$km$\,$s$^{-1}$. The system
temperature was of $\sim$1000$\,$K. All the line intensities were calibrated 
in antenna temperature (T$_A^*$). 

Fig.$\,$1 shows the emission of H$^{13}$CO$^{+}$,
HN$^{13}$C and of shock tracers like SiO, CH$_3$OH, SO, CS, H$_2$S,
SO$_2$, and OCS, measured toward the 5--point map of L1448--mm.
The line profiles, except those of SiO, have
line widths of $\sim$1$\,$km$\,$s$^{-1}$ and are double peaked showing the 
preshock component at 4.7$\,$km$\,$s$^{-1}$ and the shock--precursor component
at 5.2$\,$km$\,$s$^{-1}$ (vertical dotted lines; Jim\'enez--Serra et
al. 2004). Toward L1448--mm
(0,--20), SiO, CH$_3$OH, CS, and SO also show broader 
redshifted emission with terminal velocities that vary from 
$\sim$7.5$\,$km$\,$s$^{-1}$ in the case of CS, to 
$\sim$14$\,$km$\,$s$^{-1}$ for CH$_3$OH (see dashed horizontal line 
in Fig.$\,$1). To estimate the molecular column densities, we have 
assumed optically thin emission. For the preshock and shock--precursor 
components we used excitacion temperatures (derived from the 
methanol emission) of $\sim$7--14$\,$K for all positions except toward
L1448--mm (0,--20), where the derived excitation
temperature was of $\sim$22$\,$K. For the postshock gas, we obtained excitation
temperatures of $\sim$20--45$\,$K.
To derive the fractional abundances, the H$_2$ column density for 
the preshock and shock--precursor components was estimated from the 
H$^{13}$CO$^{+}$ column densities, while for the postshock gas, 
we used the broad CO {\it J}=1$\rightarrow$0 emission. 
We considered CO and HCO$^{+}$ fractional
abundances of $\sim$10$^{-4}$ and $\sim$10$^{-8}$ respectively 
(Irvine, Goldsmith, \& Hjalmarson 1987), and a $^{12}$C$/$$^{13}$C ratio of 90. 
In Fig.$\,$2, we show the comparison of the 
abundances of the shock tracers derived for the
preshock, shock--precursor and postshock components
toward the L1448--mm (0,--20) position. In the following, 
we discuss the abundances of these molecules for each component.

\subsection{Preshock Gas}

Neutral species like HN$^{13}$C and HCO trace the quiescent gas
of the preshock component at 4.7$\,$km$\,$s$^{-1}$
(Jim\'enez--Serra et al. 2004). We do not detect 
SiO, OCS, or SO$_2$ at this component, which supports this idea. 
The molecular emission observed toward L1448--mm (0,20) and (20,0) peaks at 
4.7$\,$km$\,$s$^{-1}$, indicating that these regions are unaffected 
by the precursor (Jim\'enez--Serra et al. 2004). The 
upper limits to the SiO abundance of $\leq$8$\times$10$^{-13}$
and to the OCS and SO$_2$ abundances of $\leq$10$^{-10}$,  
are respectively of the same order of magnitude
and one order of magnitude smaller than those found in 
L134N or TMC--1 (Ziurys et al. 1989; Mart\'{\i}n--Pintado
et al. 1992; Matthews et al. 1987; Ohishi, Irvine, \& Kaifu 1992). 
The CH$_3$OH, SO, CS, and H$_2$S abundances are similar to those derived 
from dark clouds ($\sim$10$^{-10}$--10$^{-9}$; Ohishi et
al. 1992; Mart\'{\i}n--Pintado et al. 1992).


\begin{figure}
\epsscale{0.99}
\plotone{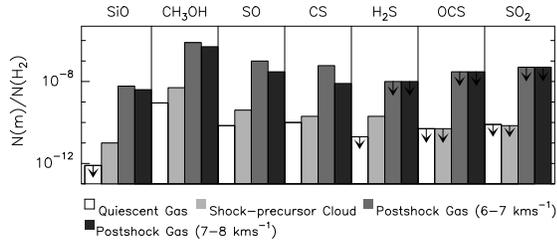}
\caption{Derived abundances of SiO, CH$_3$OH, SO, CS, H$_2$S, OCS, and SO$_2$
  for the preshock (white), the shock--precursor (light grey),
  and the postshock components (grey and dark grey) toward L1448--mm
  (0,--20). The black arrows indicate the upper limits to the molecular
  abundances estimated from the 3$\sigma$ level noise
  in the spectra. SiO, CH$_3$OH, SO, and CS are enhanced by up to a factor of
  10 in the shock--precursor component and by a factor of $\sim$1000
  in the postshock gas. As H$_2$S is only enhanced in the
  shock--precursor component, and OCS and SO$_2$ are undetected, 
  SO and CS are the most abundant 
  sulfur--bearing molecules in the shocked gas.}
\label{fig2}
\end{figure}


\subsection{Shock--precursor Component}

The very narrow ($\sim$0.6$\,$km$\,$s$^{-1}$) SiO emission
detected at 5.2$\,$km$\,$s$^{-1}$ with an abundance of $\sim$10$^{-11}$,
and the enhancement of the molecular ions 
like H$^{13}$CO$^{+}$ and N$_2$H$^{+}$,
have been proposed as signatures of the interaction of the C--shock
precursor (Jim\'enez--Serra et al. 2004). Like the ions, our data show that 
CH$_3$OH, H$_2$S, SO, and CS peak at the shock--precursor 
component in the regions where we detect narrow SiO (see Fig.$\,$1). This
is even more clear toward L1448--mm (0,--20), 
where the velocity peaks of these molecules are slightly
redshifted ($\sim$5.4--5.6$\,$km$\,$s$^{-1}$) just in the same direction
as the broader postshock gas, suggesting that 
the material is already accelerated by the precursor. 
Like SiO, the abundances of CH$_3$OH, H$_2$S, SO, and CS
are enhanced by up to a factor of 10 in the shock--precursor 
component with respect to the quiescent gas (see Fig.$\,$2).
Assuming that grain mantles are completely removed in the postshock regime (see
section 2.3), this enhancement implies the ejection 
of the 0.1\% of the total mantle material in the shock precursor. 
SO is only detected toward L1448--mm (0,0) and (0,--20), 
showing a similar behavior to SiO (Fig.$\,$1). 
The upper limits to the SO abundance toward L1448--mm (--20,0), (0,20),
and (20,0) are of $\leq$10$^{-10}$. All these facts suggest that 
the production of these molecules is closely related to the precursor
interaction. OCS and SO$_2$ are undetected also at this
component (see Fig.$\,$2). 

\subsection{Postshock Gas}

L1448--mm (0,--20) is the only position where 
SiO shows broad emission suggesting
that the gas has already entered the shock
($v_s$$\sim$10$\,$km$\,$s$^{-1}$; Jim\'enez--Serra et al. 2004). 
As expected for CH$_3$OH, SO, and CS, these molecules also
present broad line profiles toward this position. 
The SiO and CH$_3$OH abundances enhanced by a factor of $\sim$1000 
in the postshock gas ($\sim$10$^{-8}$ and 10$^{-6}$ respectively;
Fig.$\,$2), are similar to those observed in other outflows 
(Bachiller \& P\'erez Guti\'errez 1997; Blake et al. 1987), 
suggesting the idea that icy mantles have been completely removed from grains 
(nearly 90\% of the mantle material
is injected into gas phase for $v_s$$\leq$15$\,$km$\,$s$^{-1}$;
Caselli, Hartquist, \& Havnes 1997). The enhancement by also three orders of
magnitude of SO and CS in the postshock regime (Fig.$\,$2) 
clearly contrasts with the lack of broad H$_2$S
emission toward L1448--mm (0,--20) (Fig.$\,$1). 
The upper limits to the H$_2$S abundance of $\leq$10$^{-8}$ 
are a factor of 10 smaller than 
those observed in evolved outflows
($\sim$10$^{-7}$; Blake et al. 1987; 
Bachiller \& P\'erez Guti\'errez 1997) and 
predicted by sulfur chemistry models with injection
of H$_2$S from grains (Charnley 1997).
Since we do not either detect OCS or SO$_2$
arising from the postshock gas (Fig.$\,$2), 
SO and CS are the most abundant sulfur--bearing molecules at the
first stages of the shock evolution toward L1448--mm. 
The molecular abundances are not significantly
altered by the ejection of CO from grains since  
its fractional abundance remains unchanged
after mantle desorption (Markwick et al. 2000).

\section{Discussion}

The observed trend in the shock tracers to be enhanced in the shock--precursor
and the postshock components (Fig.$\,$2), supports an 
evolutionary effect produced by the propagation of the shock into the
ambient gas. The material affected by the young shocks in L1448--mm has
not reached the postshock chemical equilibrium \citep{flo99}. One then
expects to distinguish between the molecules ejected
from grains and the species generated in gas phase. 
This will help us to understand the peculiar grain chemistry
observed toward L1448--mm, specially in the case of the sulfur chemistry,
which seems to be strongly dependent on the chemical ``history''
of the grain mantles in this region.

The abundances of shock tracers like SiO, 
CH$_3$OH, SO, CS or H$_2$S in the preshock gas are similar to those found
in dark clouds. As the shock aproaches the unperturbed gas, the first
interaction will be driven by the precursor,
modifying the physical conditions and the chemistry of the ambient cloud
(Draine \& McKee 1993). Jim\'enez--Serra et al. (2004) proposed that
the enhancement of the ions in the shock--precursor
component toward L1448--mm, is explained by the interaction of
the magnetic precursor that forces the
ions to slip \citep{dra80} from the neutral quiescent gas at
4.7$\,$km$\,$s$^{-1}$ to the shock--precursor component at
5.2$\,$km$\,$s$^{-1}$. This 
ion--neutral velocity drift produces the ejection of material from grains
\citep{flo96,mark00} enhancing the SiO, CH$_3$OH, SO, CS, and
H$_2$S abundances by up to a factor of 10 in the shock--precursor 
component (Fig.$\,$2). The detection of slightly redshifted emission 
for these shock tracers toward the L1448--mm (0,--20) position
(Fig.$\,$1), is also consistent with the kinematical effects expected for   
the precursor \citep{dra80}.
 
As the gas enters the shock and reaches the shock velocity, the rate
of molecules ejected from grain mantles is largely
increased, leading to the broadening of the line profiles of SiO,
CH$_3$OH, SO, and CS toward the L1448--mm (0,--20) position (see
Fig.$\,$1). These shock tracers are $''$instantaneously$''$ enhanced
by a factor of $\sim$1000 in the postshock gas. For
SiO and CH$_3$OH, this enhancement is consistent with the abundances
observed in other outflows (Bachiller \& P\'erez Guti\'errez 1997;
Blake et al. 1987) 
and predicted by chemistry models with injection of material from grains
(Flower et al. 1996; Charnley et al. 1992).
However, in the case of the sulfur chemistry,
the lack of broad H$_2$S emission and the fact that the time--scales
for H$_2$S to decrease its abundance to $\leq$10$^{-8}$ are much
larger (t$\geq$10$^{4}$$\,$yr; Charnley 1997) than the dynamical age 
of the outflow, contrast with the general
assumption that this molecule is also released from grains (Charnley
1997). The H$_2$S abundances derived from the shock--precursor
($\sim$2$\times$10$^{-10}$) and the postshock ($\leq$10$^{-8}$)
components cannot be responsible for the large SO and CS abundances 
($\sim$10$^{-7}$) observed in the postshock gas. 

OCS has been detected on grains (Palumbo, 
Geballe, \& Tielens 1997), and van der Tak et al. (2003) and
Mart\'{\i}n et al. (2004) have reported large OCS abundances in massive 
star--forming regions and in the nucleus of NGC$\,$253. One may
consider that OCS could play an important role in
the sulfur chemistry of L1448--mm. However, the upper limits to the
OCS abundance ($\leq$3$\times$10$^{-8}$) observed toward L1448--mm
cannot either explain the
large SO and CS abundances. The fact that SO$_2$ is also
undetected toward this young outflow suggests that OCS and SO$_2$ are
late--time products of the sulfur chemistry 
(Charnley 1997).

Finally, Wakelam et al. (2004) have recently proposed that
the sulfur of grain mantles could be directly released into gas
phase in atomic form,
or in molecules that would be rapidly converted into it. 
Recent Spitzer observations have shown that S is very
abundant toward the young Cepheus E outflow (Noriega--Crespo et al. 2004).
As noted by Wakelam et al. (2004), S or S$_2$ may produce large abundances of
SO by reacting with O$_2$ or O for t$\sim$100--1000$\,$yr. However,
the predicted SO$/$CS abundance ratio is four orders of magnitude larger than
the SO$/$CS ratio observed in the postshock gas, and
O$_2$ is undetected toward young outflows like NGC$\,$2071 or 
IRAS$\,$16293 \citep{pag03}. It is then unlikely that SO and CS have been 
formed by gas phase chemistry from atomic sulfur 
in the postshock gas of L1448--mm. 

The most likely explanation is that the
amount of SO and CS within the grain mantles of L1448--mm 
must be larger than initially thought, suggesting the
idea that these molecules could have been generated by gas phase chemistry 
in the past, and afterwards depleted onto grains (Bergin, Melnick, \& Neufeld
1998). This would be consistent with the fact that SO or CS 
are more abundant than H$_2$S or OCS in the grain mantles of the 
L1448--mm outflow. In general,
the chemistry of sulfur--bearing molecules in shocks and hot cores
will strongly depend on the ``history'' of the formation of the
grain mantles. 

In summary, the line profiles and the abundances of the
shock tracers measured at the different velocity
components in the L1448--mm outflow,
indicate that we have observed the ``finger prints'' 
of the three stages of the shock 
evolution: the preshock, the shock--precursor and
the postshock gas. The observed enhancement in the
abundance of SiO, CH$_3$OH, SO, and CS 
by one order of magnitude in the shock--precursor component,
and by three orders of magnitude in the postshock gas, indicate 
the recent ejection of these molecules from grains. The
abundances of the proposed sulfur--bearing parents like
H$_2$S and OCS, cannot explain the large enhancement of SO and CS
in the postshock gas. This suggests a strong dependence of
the sulfur chemistry on the chemical evolution of the 
grain mantle composition.
The lack of OCS and SO$_2$ in L1448--mm indicates that these
molecules must be late--time products of the sulfur chemistry. 

\acknowledgments

This work has been supported by the Spanish MEC under projects
number AYA2002--10113--E, AYA2003--02785--E and ESP2004--00665. We
are grateful to the IRAM 30$\,$m telescope staff for the help provided 
during the observations. We also thank an anonymous referee for their
useful comments and suggestions.


\begin{thebibliography}{}

\bibitem[Bachiller \& P\'erez Guti\'errez 1997]{bac97}
Bachiller, R., \& P\'erez--Guti\'errez, M. 1997, \apj, 487, L93

\bibitem[Bergin et al. 1998]{ber98}
Bergin, E. A., Melnick, G. J., \& Neufeld, D. A. 1998, \apj, 499, 777

\bibitem[Blake et al. 1987]{bla87}
Blake, G. A., Sutton, E. C., Masson, C. R., \& Phillips, T. G. 1987,
\apj, 315, 621


\bibitem[Caselli et al. 1997]{cas97}
Caselli, P., Hartquist, T. W., \& Havnes, O. 1997, \aap,
322, 296

\bibitem[Charnley et al. 1992]{cha92}
Charnley, S. B., Tielens, A. G. G. M., \& Millar, T. J. 1992, \apj,
399, L71

\bibitem[Charnley 1997]{cha97}
Charnley, S. B. 1997, \apj, 481, 396


\bibitem[Draine 1980]{dra80}
Draine, B. T. 1980, \apj, 241, 1021

\bibitem[Draine \& McKee 1993]{dra93}
Draine, B. T., \& McKee, C. F. 1993, \araa, 31, 373

\bibitem[Duley et al. 1980]{dul80}
Duley, W. W., Millar, T. J., \& Williams, D. A. 1980, \mnras, 192, 945


\bibitem[Flower et al. 1996]{flo96}
Flower, D. R., Pineau des For\^ets, G., Field, D., \& May, P. W. 1996,
\mnras, 280, 447

\bibitem[Flower \& Pineau des For\^ets 1999]{flo99}
Flower, D. R., \& Pineau des For\^ets, G. 1999, \mnras, 308, 271



\bibitem[Irvine, Goldsmith \& Hjalmarson 1987]{irv87}
Irvine, W. M., Goldsmith, P. F., \& Hjalmarson, \AA. 1987,
in Interstellar Processes, ed. D. J. Hollenbach, \&  H. A. Thronson
(Dordrecht: Reidel), p. 561

\bibitem[Jim\'enez--Serra et al. 2004]{jim04}
Jim\'enez--Serra, I., Mart\'{\i}n--Pintado, J.,
Rodr\'{\i}guez--Franco, A., \& Marcelino, N. 2004, \apj, 603, L49

\bibitem[Markwick et al. 2000]{mark00}
Markwick, A. J., Millar, T. J., \& Charnley, S. B. 2000, \apj, 535, 256

\bibitem[Mart\'{\i}n et al. 2004]{smar04}
Mart\'{\i}n, S., Mart\'{\i}n--Pintado, J., Mauersberger, R.,
Henkel, C., \& Garc\'{\i}a--Burillo, S. 2004, \apj, 620, 210 

\bibitem[Mart\'{\i}n--Pintado et al. 1992]{mar92}
Mart\'{\i}n--Pintado, J., Bachiller, R., \& Fuente, A. 1992, \aap,
254, 315

\bibitem[Matthews et al. 1987]{mat87}
Matthews, H. E., MacLeod, J. M., Broten, N. W., Madden, S. C., \&
Friberg, P. 1987, \apj, 315, 646


\bibitem[Noriega--Crespo et al. 2004]{nor04}
Noriega--Crespo, A., Moro--Mart\'{\i}n, A., Carey, S., Morris, P. W.,
Padgett, D. L., Latter, W. B., \& Muzerolle, J. 2004, \apjs, 154, 402

\bibitem[Ohishi et al. 1992]{ohi92}
Ohishi, M., Irvine, W. M., \& Kaifu, N. 1992, in IAU Symp. 150, 
Astrochemistry of Cosmic Phenomena, ed. P. D. Singh (Dordretch:
Kluwer), p. 171

\bibitem[Pagani et al. 2003]{pag03}
Pagani, L., et al. 2003, \aap, 402, L77

\bibitem[Palumbo, Geballe \& Tielens 1997]{pal97}
Palumbo, M. E., Geballe, T. R., \& Tielens, A. G. G. M. 1997, \apj,
479, 839





\bibitem[Smith 1991]{smi91}
Smith, R. G. 1991, \mnras, 249, 172

\bibitem[Tielens \& Allamandola 1987]{tie87}
Tielens, A. G. G. M., \& Allamandola, L. J. 1987, in Physical
Processes in Interstellar Clouds, ed. G. E. Morfill, \& M. Scholer
(Dordretch: Reidel), p. 333

\bibitem[van der Tak et al. 2003]{vdt03}
van der Tak, F. F. S., Boonman, A. M. S., Braakman, R., \& van Dishoeck,
E. F. 2003, \aap, 412, 133

\bibitem[Wakelam et al. 2004]{wak04}
Wakelam, V., Caselli, P., Ceccarelli, C., Herbst, E., \& Castets,
A. 2004, \aap, 422, 159

\bibitem[Ziurys et al. 1989]{ziu89}
Ziurys, L. M., Friberg, P., \& Irvine, W. M. 1989, \apj, 343, 201


\end{thebibliography}
\end{document}